# Label-free visualization of amyloid plaques in Alzheimer's disease with polarization-sensitive photoacoustic Mueller matrix tomography

Zhenhui Zhang[1, 2, *], Yujiao Shi[1, 2, *], Qi Shen[1, 2], Zhixiong Wang[1, 2, 3], Da Xing[1, 2, †], Sihua Yang[1, 2, †]

1  MOE Key Laboratory of Laser Life Science & Institute of Laser Life Science, College of Biophotonics, South China Normal University, Guangzhou 510631, China.

2  Guangdong Provincial Key Laboratory of Laser Life Science, College of Biophotonics, South China Normal University, Guangzhou 510631, China.

3  Laboratory of Cellular Imaging and Macromolecular Biophysics, National Institute of Biomedical Imaging and Bioengineering, National Institutes of Health, Bethesda, Maryland 20892, USA

[*]These authors contributed equally to this work.

[†]Corresponding author. Email: xingda@scnu.edu.cn; yangsh@scnu.edu.cn

**Abstract**

The formation of amyloid plaques in the cortical and hippocampal brain regions caused by abnormal deposition of extracellular amyloid β-protein (Aβ) is a characteristic pathological hallmark of early Alzheimer's disease (AD), while label-free graphic rendering of diseased amyloid plaques in vivo is still a highly challenging task. Herein, by ingeniously extracting the polarization-sensitive optical absorption of amyloid plaques via photoacoustic (PA) technique, a novel PA Mueller matrix (PAMM) tomography that capable of providing three new conformational parameters of molecules is developed to realize depth-resolved label-free imaging of amyloid plaques. Whole brain PAMM imaging on different stages of APP/PS1 transgenic AD mice has been performed to demonstrate its ability for *in situ/in vivo* quantitative three-dimensional (3D) detection of amyloid plaques and its great potential for monitoring early AD pathological development without labeling.

**Introduction**

Alzheimer's disease (AD) is the most common, chronic progressive and multifaceted neurodegenerative disorder characterized by progressive memory decline and subsequent loss of broader cognitive functions *(1-3)*. By 2019, more than 50 million people in the world are living with dementia, and this figure will increase to 152 million by 2050, of which 2/3 are AD. At present, the annual cost of dementia is estimated at US $1 trillion, which will be

double by 2030 *(4)*. Unfortunately, as AD is a heterogeneous disease with a complex and multifaceted pathobiology, the identification and validation of effective biomarkers for early diagnosis and the development of disease-modifying therapies has been severely hampered *(1,5)*. Even though there is no approved diagnosis or cure of AD, the amyloid hypothesis is widely accepted to explain the pathophysiology and clinical evolution of AD *(1,6,7)*. Therefore, the amyloid plaques formed by the folding and deposition of amyloid β-protein (Aβ) in cerebral cortex and hippocampus are ideal biomarkers for early diagnosis of AD *(8-10)*. Currently, positron emission tomography (PET) and functional magnetic resonance imaging (fMRI) are recommended for the diagnosis of AD in clinic *(1,11)*. PET has been used as the most direct imaging technique to display the pathological changes of AD in vivo by providing quantitative distribution of the amyloid plaques in brain *(1,12)*. However, PET involves radioactive tracers and expensive instruments, where the affordability problem hinders its wide application in AD diagnosis1. fMRI is used to provide memory formation, cognitive changes over time, and neural activity information for the diagnosis of early AD via detecting changes in blood oxygenation levels caused by neural activity *(1,13-15)*, which lacks detection specificity of amyloid plaques and is limited by insufficient spatial resolution. The technologies of Infrared (IR) and Raman spectroscopy label-free imaging of amyloid plaques have been developed in recent years, but poor penetration depth limits their *in situ* detection capability *(16-18)*. It is still a great challenge for current medical imaging technology to realize label-free and non-destructive detection of amyloid plaques in situ and to achieve early diagnosis of AD.

As an emerging technology in biomedical field, photoacoustic (PA) imaging uniquely integrates the merit of high-contrast optical imaging and the deep-resolved ultrasound imaging, capable of providing 3D optical absorption distribution of in vivo tissues with high resolution *(19,20)*. Recently, a novel polarimetric PA imaging emerged and began becoming one of the research focuses in PA imaging *(21-26)*. The salient polarization response of Aβ is one of its optical characteristics, which forged the wide applications of polarized optical techniques, such as Mueller matrix imaging (MMI) *(27-29)*, for the detection of amyloid plaques and AD diagnosis. Unfortunately, as the MMI treats the tissue block with a certain thickness as a two-dimensional optical transmission matrix, prominent difficulty exists in depth-resolved interpreting the highly-mixing intrinsic polarimetry properties with physically meaningful and explicit information *(30,31)*. Even worse, the MMI is highly sensitive to the thickness of the prepared tissue samples for both transmission and reflection imaging modes, which severely limits the application of this technique for in vivo clinical applications *(32)*. Novel imaging technique that allows for depth-resolved label-free identifying and quantification distribution of amyloid plaques in the brain with physically explicit molecular conformations-sensitive parameters is urgently needed.

The previous studies have been demonstrated that PA technology has a unique advantage in whole brain imaging because of its features of real-time, low-cost, high-resolution, and ease of incorporation into portable devices *(33)*, and visualizing the biodistribution of amyloid-β (Aβ) deposits and abnormal tau accumulation within mouse models of AD amyloidosis by PA technique has been realized. However, the reported works heavily rely on probe labeling *(34-38)*, owing to the low optical specificity of the amyloid plaques at tissue optical window. When goes into real applications, the probe labeling will bring serious biosafety risks, not to mention that the detection accuracy highly depends on the unexpectable labeling efficiency. Herein, by employing the state-of-the-art 3D absorption imaging capability of the PA technique, a novel polarization-sensitive PA Mueller matrix (PAMM) tomography technique has been theoretically proposed and experimentally demonstrated, to render quantitative information of spatial distributions for optically-anisotropic molecules and tissues via ingeniously detecting the molecules' characteristic absorption diversity under different polarized light excitation. Three new parameters, the degree order of tissue spatial arrangement ranging 0 to 1, orientation ranging 0° to 180° and circular dichroism has been proposed and experimentally verified by phantom and tissue samples. The brain model of APP/PS1 transgenic mice (AD mice) has been established and imaged by the PAMM technique to demonstrate its ability for label-free *in situ/in vivo* 3D AD detection by recognizing polarization-sensitive amyloid plaques, and quantified the accumulation changes of amyloid plaques in AD mice at different stages. The PAMM technique allows label-free imaging of amyloid plaques distribution in the whole brain cortex and hippocampus of mice, which provides a powerful imaging technology for studying the pathological development of AD disease *in vivo* and evaluating the effect of preclinical drug treatment.

**Results**

**Principles of the PAMM tomography.** When the incident light irradiates a molecule or a chromophore, the quantum state transition process from the ground state $|\varphi_a\rangle$ to the excited state $|\varphi_b\rangle$ is a resonance coupling between the electromagnetic wave and the electric and magnetic dipole transition moment of molecule, and the spatially fixed transition dipole moments (TDMs) is the combination of the electric transition dipole moment (ETDM) $\mathbf{U}^{ab} = \langle \varphi_a | \hat{u} | \varphi_b \rangle$, and magnetic transition dipole moment (MTDM) $\mathbf{M}^{ab} = \langle \varphi_a | \hat{m} | \varphi_b \rangle$, with $\hat{u}$ and $\hat{m}$ to be the electric and magnetic dipole moment operators, respectively *(39-41)*. In order to obtain the optical absorption, the Hamiltonian operator of energy interaction between a molecule and the electromagnetic waves with electric, **E**, and magnetic, **B**, fields are introduced by reasonably neglecting higher-order interaction terms as $\hat{H}^{int} = -\hat{u} \cdot \mathbf{E} - \hat{m} \cdot \mathbf{B}$ *(40)*. Therefore, according to the Fermi Golden Rule in quantum mechanics, the intensity of optical absorption from $|\varphi_a\rangle$ to $|\varphi_b\rangle$ can be written as follows *(40)*

$$w_\alpha(|\varphi_a\rangle \to |\varphi_b\rangle) = \gamma \left( \mathbf{U}^{ab*} \cdot \hat{\mathbf{E}}^* \mathbf{U}^{ab} \cdot \hat{\mathbf{E}} + \mathbf{M}^{ab*} \cdot \hat{\mathbf{B}}^* \mathbf{M}^{ab} \cdot \hat{\mathbf{B}} + \mathbf{U}^{ab*} \cdot \hat{\mathbf{E}}^* \mathbf{M}^{ab} \cdot \hat{\mathbf{B}} + \mathbf{U}^{ab} \cdot \hat{\mathbf{E}} \mathbf{M}^{ab*} \cdot \hat{\mathbf{B}}^* \right). \quad (1)$$

Here, $\hat{\mathbf{E}} = \mathbf{E} \cdot \hat{e}$, $\hat{\mathbf{B}} = \mathbf{B} \cdot \hat{b}$, and $\hat{e}$, $\hat{b}$ are the unit vector along **E**, **B** respectively. γ is a constant for a certain molecule.

Basically, the optical absorption is mainly contributed by their ETDMs owing to the negligible magnetic dipole moments under linearly polarized light irradiation for most non-helical molecules or chromophores without chirality *(42)*. For this case, the second, third and fourth terms on the right of Eq. (1) become zero. Therefore, when a chromophore is irradiated by a linearly polarized light with a certain included angle *θ* to the ETDM, as indicated the inset in upper-right corner of Fig. 1A, the optical absorption thus reduces to a *θ* dependent function as

$$w_\alpha(\theta) = \gamma |\mathbf{U}^{ab} \cdot \mathbf{E}|^2 = \gamma |\mathbf{U}^{ab}|^2 \cdot \mathbf{E}^2 \cos^2(\theta) \quad (2)$$

Eq. (2) indicates an obvious difference of the optical absorption when the angle between the ETDM and the light polarization orientation changes. As demonstrated in the left part of Fig. 1B, when the polarization orientation of the incident light is coinciding with the ETDM orientation, i.e., $\theta=0$, the transition probability or optical absorption of the molecule exhibits a maximum value *(39,40)*. While for the opposite case that the light polarization orientation is perpendicular to the ETDM orientation, a minimum value of optical absorption then can be observed *(39,40)*. Meanwhile, for the case that helical molecules or chromophores with chirality are irradiated by circularly polarized light, the MTDMs are no more negligible, where the interference of the ETDMs with MTDMs will induce a rotatory strength for the transition as $R_0 = I_m(\langle \varphi_a|\hat{u}|\varphi_b\rangle \cdot \langle \varphi_b|\hat{m}|\varphi_a\rangle)$, and optical transition changes with the light polarization *(43-45)*. The circular dichroism (CD) referring as the absorption difference of left and right-handed circularly polarized lights then can be obtained according to Eq. (1) by inducing the left and right polarization vector $\hat{e}_L$ and $\hat{e}_R$ as *(40)*

$$w_{\alpha\text{-CD}} = 2\gamma \left( \mathbf{U}^{ab*} \cdot \hat{\mathbf{E}}_R \mathbf{M}^{ab} \cdot \hat{\mathbf{B}}_R^* - \mathbf{U}^{ab*} \cdot \hat{\mathbf{E}}_L \mathbf{M}^{ab} \cdot \hat{\mathbf{B}}_L^* \right) \tag{3}$$

where, $\hat{\mathbf{E}}_{L,R} = \mathbf{E} \cdot \hat{\mathbf{e}}_{L,R}$ and $\hat{\mathbf{B}}_{L,R} = \mathbf{B} \cdot \hat{\mathbf{b}}_{L,R}$. As indicated by Eq. (3) and illustrated in the right part of Fig. 1B, for chiral molecules or chromophores such as protein secondary structure and DNA, obvious optical absorption differences can be observed between left- and right-handed circularly polarized light excitation *(40)*.

Benefiting from the unique 3D optical absorption mapping capability of PA imaging, the authors creatively embedded the polarized-light-excited absorption transition into the PA technique, thus capable of rendering characteristic information of the optically-anisotropic molecules via ingeniously detecting their absorption variation under polarized laser excitation, where according to PA theory, the PA signal intensity can be written as

$$P = \eta_{th} I_0 \Gamma \mu_\alpha \left( |\varphi_a\rangle \to |\varphi_b\rangle \right). \tag{4}$$

Here, $\eta_{th}$, $\Gamma$ and $I_0$ are the heat conversion efficiency, the Grueneisen parameter, and the optical fluence. $\mu_\alpha \left( |\varphi_a\rangle \to |\varphi_b\rangle \right)$ is the optical absorption coefficient (see methods).

To achieve this goal, a state-of-the-art PAMM tomography is established in analogy to the optical MMI by treating the sample as a 3D optical absorption matrix. Different from the optical MMI that treats the sample block as a two-dimensional optical transmission matrix, the proposed PAMM tomography is capable of providing depth-resolved interpretation of sample's intrinsic polarimetry properties with physically explicit information. In the PAMM imaging, six laser beams with different polarization states, 0°, 90°, 45°and -45°linearly polarized laser beams, and left-, right-handed circularly polarized laser beams, are introduced as the excitation beams, where their polarization states can be described by the Stoke vector as $\mathbf{S}_{Laser} = (S_0 \quad S_1 \quad S_2 \quad S_3)^T$. By sequentially obtaining the corresponding generated PA signal amplitudes under different polarized laser beams, the PAMM can be obtained and satisfied by the following formula

$$\mathbf{S}_{PA} = \mathbf{M}_{PA} \cdot \mathbf{S}_{Laser} \tag{5}$$

$\mathbf{S}_{PA}$ and $\mathbf{M}_{PA}$ are the resultant PA signal matrix and the defined PAMM of tissue, respectively. Results of matrix calculation show that, the first line of $\mathbf{M}_{PA}$ indicates exact absorption information of the excited molecules under different polarized laser beams as (see methods)

$$\left(M_{11}\ M_{12}\ M_{13}\ M_{14}\right) = \left(\frac{PA_H + PA_V}{2}\ \frac{PA_H - PA_V}{2}\ \frac{PA_P - PA_M}{2}\ \frac{PA_R - PA_L}{2}\right). \tag{6}$$

Here, $PA_H$, $PA_V$ are the PA signal amplitudes under horizontal (0°) and vertical (90°) linearly polarized lasers. $PA_P$, $PA_M$ are the PA signal amplitudes under 45° and -45° linearly polarized lasers. $PA_R$, $PA_L$ are the PA signal amplitudes under right- and left-handed circularly polarized lasers. According to Eq. (6), a parameter $PA = 2M_{11}$ that reflecting the total absorption coefficient of excited molecules is firstly defined.

Basically, the changes of ordered degree for a molecule or a group molecule is a reflection of changes in their interaction, structure and functions. Here, by analogizing to the definition of degree of light polarization (DLP) in optics *(24)*, a quantitative parameter, the degree of anisotropy (DOA) that reflects the ordered degree of a molecule or a group molecule by extracting the optical absorption anisotropy is defined according to the obtained PAMM as

$$DOA = \frac{\sqrt{M_{11}^2 + M_{13}^2}}{M_{11}} \tag{7}$$

which gives a quantitative value ranging from 0 to 1. In practical applications, usually a group of distributed molecules is excited in a laser spot, as indicated in the Supplementary Fig. S1A. The case DOA=1 (i.e., $\Delta\theta = 0$) implies the optical absorption to be completely anisotropic, indicating the ETDMs of all excited molecules are in a consistent arrangement. In this case, huge variation of the corresponding PA signal amplitude with the polarization angle of excitation laser beams can be observed, as shown in the Fig. 1C and Supplementary Fig. S1B. When the ordered degree of molecules changes, the averaged ETDM $\overline{\mathbf{U}^{ab}} = \sum_n \mathbf{U}_n^{ab}$ at the excited point is a vector addition of ETDMs for all molecules, with an added dispersion angle $\Delta\theta$ to the averaged ETDM that can be quantitatively defined by the DOA. For the case that the molecules are randomly distributed, $\Delta\theta = 45°$ and DOA becomes zero, where no variation of the corresponding PA signal amplitude will be observed.

Further, the orientation of the transition dipole moment (OTDM) for the excited molecules, i.e., the angle $\theta$, can also be ingeniously obtained by the matrix calculation of PAMM as

$$\text{OTDM} = \frac{1}{2} ar\tan\left(\frac{M_{13}}{M_{12}}\right) \quad (8)$$

Eq. (8) is carefully validated by simulation study of the corresponding PA signal amplitude as a function of laser polarization angle with different values of OTDM as shown in the Supplementary Fig. S1C. According to the Eq. (8), the calculated mathematical OTDM angle ranges from -45° to 45° rather than -90° to 90° (or 0° to 180°) owing to the data degeneracy arising from the data symmetric about the center of 0°, which originates from the scalar property of the PA signal amplitude. Therefore, the OTDM of the two transition dipoles moment perpendicular to each other are degenerate, shown as U1 and U2 in the I and II region of Supplementary Fig. S2A. In order to lift the degeneracy, we extend the range from -45°~-45° to -45°~135° by recognizing the quadrant of the OTDM through the relationship between $PA_V$ and $PA_H$. When $PA_V > PA_H$, the OTDM angle is defined by adding the calculated OTDM according to Eq. (8) with a 90° compensation angle. Therefore, the OTDM angle is then extended to the range of -45° ~135° (Supplementary Fig. S2C). Finally, all angles are added to 45° to achieve 0°~180° OTDM imaging.

Apart from the molecular spatial arrangement information, chirality is also one of the most basic characters for molecules in nature and life. Benefiting from the introduction of circularly polarized laser beam excitation to the PAMM tomography, the molecular CD can also be calculated through matrix calculation as

$$CD = \left|\frac{M_{14}}{M_{11}}\right| \quad (9)$$

It should be noticed that, the extracted CD information through the proposed PAMM tomography can be used for 3D imaging, which is impossible for existed method. By theoretically establishing the PAMM tomography, three quantitative parameters, DOA, OTDM, and CD that reflect the degree of anisotropy, the spatial arrangement and the chirality of molecules respectively, have been proposed, which can facilitate quantitative investigation of the correlation between physically explicit tissue polarimetry properties with molecular structures and functions, prefiguring advanced scientific significance for the analysis of pathological and physiological characteristics of biological tissues and effective diagnosis of difficult diseases. The right part of Fig. 1D schematically describes the AD process involving Aβ continuous deposition, and Aβ with vector absorption characteristics generating PA signals when irradiated by polarized light.

**Demonstration of PAMM tomography capability.** In order to demonstrate the efficiency of the proposed method, optically anisotropic phantom samples are prepared and imaged. As shown in the Fig. 2A, four dichroic polyvinyl alcohol (PVA) strips with strong anisotropic optical absorption properties were imaged, by comparing with an optically isotropic polyvinyl chloride (PVC) strip. Four linearly polarized laser beams with a polarization angle difference of 45° were applied to form the polarized PA microscopy images, as shown in the Fig. 2B. Results show that, the PA image amplitudes of the PVA samples strongly depend on the angle between the orientations of ETDMs and the excitation laser polarization direction, while negligible changes of image amplitude is observed for the optically isotropic PVC sample. Further, in order to verify the ability of the PAMM tomography for the characterization of molecular chirality, a phantom sample composed of chiral L-polylactide (L-P) and non-chiral PVC is prepared and imaged. Fig. 2C is the schematic diagram of the prepared sample. The corresponding PA images under left- and right-handed circularly polarized lasers excitation are shown in the Fig. 2D. Further, we can get Figs. 2E-H by the PAMM tomography. The PA image is shown in Fig. 2E, where image amplitudes for the four PVA samples are basically the same. After data processing, the quantitative DOA image of the optically anisotropic PVA samples were selectively obtained with a DOA value of about 0.8 (Fig. 2I). The corresponding OTDMs of the PVA samples were presented in the Fig. 2G, which coincides well with the setting orientations in Fig. 2A. Fig. 2J is the statistical diagram of PA signal amplitudes changing with OTDM. The amplitude of PA signal comes from Fig. 2B, and the value of OTDM comes from Fig. 2G. The results of Fig. 2J show that the phase delay of the fitted curve is 45°, which is in good agreement with the theoretical value. The CD image presented in the Fig. 2H obtained by the PAMM tomography indicates that the chiral sample can be clearly distinguished from the non-chiral sample. The corresponding signal amplitudes along the green line in Fig. 2D and statistical results DOA value in Fig. 2H are presented in the Fig. 2K. These results validated the PAMM theoretical model, where the quantitative DOA, OTDM and instinct chirality information of optically anisotropic samples can be experimentally obtained and imaged by the proposed PAMM tomography.

**PAMM tomography application demonstration.** The changes of ordered degree for molecules in a material can result in changes of their intermolecular interaction, structure and functions. Precision detection and recognition of these changes possesses significant values for evaluating their performance such as efficacy and lifecycles. Here, we demonstrated that, the proposed PAMM tomography has strong ability to evaluate molecular changes during material denaturation. As indicated in Fig. 3A, an optically anisotropic PVA strip is prepared and heated at one end to simulate the non-uniform environmental heating condition. After that, the sample is imaged by the PAMM tomography. As shown in the Figs. 3B-D, no obvious PA amplitude changing has been observed when overheating, but the value of DOA for the heated area exhibits sharp decrease, indicating that the ordered molecular structure has been destroyed. Similar conclusion can be drawn by the OTDM imaging in Fig. 3C, where the values of OTDM in the heated area become randomly distributed. The inset in Fig. 3D are the scanning electron microscope image (SEMI) and the variation of PA signal amplitude with angle $\theta$ in polar coordinates at the 1, 2, 3, 4 areas of the four areas on the sample. Results indicate that, with the distance to the heating center decreases, the destroy of the sample's ordered molecular structure becomes serious, which coincides well with our PAMM experiments. The corresponding statistical results of DOA value in the 1, 2, 3, 4 areas and variation of DOA value along the red dashed lines are presented in the Fig. 3E. These results indicate that, the proposed PAMM tomography can be used as an effective method for evaluating molecular changes in material field.

The imaging depth of PAMM tomography based on polarized light excitation is a significant concern for biological applications. In order to validate PAMM imaging capability in deep-seated tissue, PAMM imaging of optically anisotropic samples buried in tissue-mimicking scattering medium is performed. The scattering medium is made of 0.25% intralipid and 3% agar in distilled water, with a reduced scattering coefficient estimated to be about 0.25 mm$^{-1}$ and a transport mean free path to be $\ell$ ~4 mm *(29)*. Three PVA strips with strong anisotropic optical absorption properties and ETDMs in different directions are inserted into the scattering medium as the imaging samples (Fig. 3F). The result of 3D PAMM imaging are shown in the Fig. 3G. These results show that the imaging depth of PA, DOA and OTDM can reach up to about 8 mm, which is about two times of the transport mean free path. The signal intensity of Fig. 3G is statistically analyzed in Fig. 3H, which shows that in the one transport mean free path depth region (0~4 mm), the DOA value and

OTDM almost remains unchanged, while the PA signal amplitude suffers intense decrease with the depth. These results indicate that the PAMM imaging behaves robust 3D imaging capability in deep-seated biological tissues. Similar conclusion can be driven by the Supplementary Fig. S6.

**PAMM tomography imaging of APP/PS1 transgenic mouse (AD mouse) and non-transgenic wild-type mouse (WT mouse).** The abnormal transformation from proteins to amyloid fibrils is closely related to the conformational diseases including the common neurodegenerative disorders such as AD and prion diseases *(6)*. With the trend of worldwide population aging intensifying, AD has developed into the third largest elderly health disease besides cardiovascular disease and cancer *(46,47)*. Early diagnosis and intervention of AD possess great significance for effectively delaying the deterioration of the disease. One of the most conspicuous features in the development of AD is the Aβ deposition in the cerebral cortex and hippocampus *(6, 46,47)*. Reports show that, the amyloid plaques, including those of Aβ fibrils, consist of many intertwined protofilaments *(48,49)*, which results in their obvious linear dichroism and circular dichroism properties when interacting with polarized light *(49-53)*. Therefore, PAMM image has great potential for the AD diagnosis. Here, the amyloid plaques deposition of AD mouse in cerebral cortex and hippocampus is imaged by the PAMM imaging. During the experiment, the brain samples of 9-month-old AD mouse and WT mouse were fixed with 4% paraformaldehyde (4 °C, 12 hours), dehydrated with 15% sucrose solutions (4 °C, 12 hours), and then immersed into sucrose solutions 30% (4 °C, 12 hours). Then the samples (as indicated in Fig. 4A) were performed by the PAMM imaging system. The samples were cut into 10 μm and 500 μm thick frozen coronal brain sections (Fig. 4A), and then were subjected to immunofluorescence imaging (Figs. 4B, C) and PAMM imaging (Figs. 4D-G), respectively. Results indicated that compared with the WT group, AD mouse showed higher DOA value (Figs. 4F, G), and increasing speckles (Figs. 4D, E). Immunofluorescence images and DOA are shown in Figs. 4C, G with good correlation between the two imaging modalities. In Fig. 4H, the difference of PA amplitude and DOA values between AD and WT mice is statistically analyzed. Estimations of the DOA value yield statistically significant differences and no significant difference is observed on PA amplitude for the WT and AD mice. Further, the concentrations of Aβ 1-42 in the cortex and hippocampus were determined with an enzyme-linked immuno-sorbent assay (ELISA), as shown in Fig. 4I. Results show the concentrations of soluble and insoluble Aβ 1-42 in protein extracted from cerebral cortex and hippocampus of AD mice were 810.734 pg/mg and 854.150 pg/mg, respectively, and the concentrations of soluble and insoluble Aβ 1-42 in protein extracted from cerebral cortex and hippocampus of WT mice were 8.709 pg/mg, 5.039 pg/mg, respectively. j, Polarized absorption spectroscopy of the concentration for 200 μM synthetic Aβ 1-42 fibrils and protein in the cortex and hippocampus of AD mouse, respectively. Finally, we tested the polarized light absorption spectra of the synthesized Aβ 1-42 fibers and the proteins obtained from the cerebral cortex and hippocampus of AD and WT mice. Results in Fig. 4J show that there are clear differences in the polarization absorption of the synthesized Aβ 1-42 fibers and the brain protein of AD mice. However, the difference between the polarization absorption of WT mice is quite low (Supplementary Fig. S7). Fig. 4K is the degree of linear polarization (DLP) *(30)* of the synthesized Aβ 1-42 fibers, the AD and WT mice brain tissues. These results indicate that the Aβ in AD mice brain tissue has obvious characteristics of polarization absorption, and therefore the DOA value can be used to evaluate protein deposition in the brain of mice. These results demonstrated that the proposed PAMM method is capable of imaging amyloid plaques in fresh brain tissues without any exogenous labels.

Furthermore, the whole brain of AD mice and WT mice were imaged by 3D PAMM in situ. The imaging results are shown in Fig. 5. Compared with the imaging results of Figs. 5A-C, it can be found that the DOA imaging results show larger DOA values in cerebral cortex and hippocampus of AD mouse than that of WT mouse (as shown in the white dotted line). The overlay images (Figs. 5D, E) are constructed from of the combination of PA imaging and DOA imaging. It can be seen that the DOA values in the cerebral cortex and hippocampus of AD mouse at different depths are higher. Compared with the immunofluorescence results (Supplementary Fig. S9), these areas correspond to the deposition areas of Aβ proteins. Further, we performed in situ deep-DOA imaging on the brains of AD and WT mice at different stages (2, 5, 10-month-old), and quantitatively analyzed the conformational changes of brain molecules at different stages using polarization sensitive DOA values. In the experiment, the whole brain tissue of mice was fixed and dehydrated, then directly imaged by polarized laser which irradiated from the horizontal upper surface of brain tissue (as shown in Supplementary Fig. S10). The horizontal section imaging results of DOA images are selected at a depth of about 1.2 mm, as shown in Fig. 5F. It can be seen that no significant difference between the 2-month-old AD and WT mice is observed. However, there was a significant difference between the 2, 5, 10-month-old AD mice, and the value of the DOA data increased with age (Fig. 5G). The experimental results show that PAMM imaging can identify the changes of brain structure in AD mice of different ages, which provides a feasible label-free, deep-imaging method to identify the development of AD.

Finally, we performed in vivo PPAM tomography brain imaging of 8-month-old WT and AD mice. The mice were firstly anesthetized with a small animal ventilator, as shown in the Fig. 6A, and corresponding DOA images overlay with PA results were shown in Figs. 6B, C. Obviously, the dispersed distribution of large DOA value regions are observed in the brain cortex of AD mouse, indicating the accumulations of amyloid plaques in the AD mouse brain. The statistical analysis results of PA amplitude and DOA value are presented in Fig. 6D. The DOA imaging results coincide well with the corresponding immunofluorescence results presented in the Figs. 6E, F. The high sensitivity of PAMM imaging for microstructure or molecular structure shows great potential for studying the pathological development of early AD disease.

**Discussion and conclusions**

Different from those polarized optical techniques that utilize transmitted or backscattered photons as detection signals, the proposed PAMM tomography uses acoustic waves produced by absorbed photons, which can inherit the depth-resolve and high-resolution characteristics of PA imaging with extended imaging depth. Nevertheless, we should notice that, the effective imaging depth of the proposed method is much beyond the ballistic regime, but it is still limited by the depolarization behavior of the polarized photons in light transmission pathway. Additionally, PA imaging technology has been applied in human brain imaging *in vivo* (54), which indicates that the presented PAMM imaging approach has strong potential to be translated to clinical diagnosis of early AD.

In conclusion, by ingeniously detecting the molecules' characteristic absorption variation under different polarized light excitation via the PA technique, we have proposed a PAMM technique with the capabilities of rendering 3D and multi-parameter quantitative spatial microstructure information about optically-anisotropic of molecules and tissues without labeling. We established the theoretical PAMM model based on the physical model of anisotropic absorption of molecules, and three newly proposed parameters, the DOA, the OTDM and the CD were obtained. We further verified the feasibility of the proposed method by optically anisotropic phantom samples and in vitro biological tissues. The 3D and label-free identification of Aβ protein deposition in a mouse model of AD demonstrates that the proposed imaging method can be used to characterize the development of AD pathology and to evaluate preclinical drug therapy.

**Methods**

**Some details of PAMM model.**

We define $\mu_\alpha(|\varphi_a\rangle \to |\varphi_b\rangle)$ as the absorption coefficient, which satisfies

$$W_\alpha = \mu_\alpha(|\varphi_a\rangle \to |\varphi_b\rangle) \cdot I(\mathbf{E},\mathbf{B}). \tag{10}$$

By considering the interaction of linearly polarized light with uniaxial molecules, $\mu_\alpha(|\varphi_a\rangle \to |\varphi_b\rangle)$ can be expressed as

$$\mu_\alpha(|\varphi_a\rangle \to \langle\varphi_b|) = \gamma |\overline{\mathbf{U}^{ab}}|^2. \tag{11}$$

In order to study the properties of molecular linear dichroism, we define parallel and vertical transition dipole moment vectors i and j respectively. Then, we obtained

$$\overline{\mathbf{U}^{ab}_{//}} = \sum_n \mathbf{U_n} \cdot \mathbf{i} = \sum_n U_n \cdot \cos(\Delta\theta), \tag{12}$$

$$\overline{\mathbf{U}^{ab}_{\perp}} = \sum_n \mathbf{U_n} \cdot \mathbf{j} = \sum_n U_n \cdot \sin(\Delta\theta). \tag{13}$$

Here, we use $\mu_{//}$, $\mu_{\perp}$ to represent the optical absorption coefficient in parallel direction and vertical direction to the optic axis, respectively. Therefore, $\mu_{//}$, $\mu_{\perp}$ can be expressed as

$$\mu_{//} = \sum_n U_n^2 \cos^2(\Delta\theta), \tag{14}$$

$$\mu_{\perp} = \sum_n U_n^2 \sin^2(\Delta\theta). \tag{15}$$

Eq. (4) then can be expressed as *(29,30)*

$$PA(\theta) = \eta_{th} I_0 \Gamma \left[ \frac{\mu_{//} + \mu_{\perp}}{2} + \frac{\mu_{//} - \mu_{\perp}}{2} \cos(2\theta) \right]. \tag{16}$$

According to the definition of Stoke vector matrix *(20)*, the Stoke vector matrix of the six polarized excitation light can be expressed as

$$\mathbf{S}_{laser} = \begin{pmatrix} 1 & 1 & 1 & 1 & 1 & 1 \\ 1 & -1 & 0 & 0 & 0 & 0 \\ 0 & 0 & 1 & -1 & 0 & 0 \\ 0 & 0 & 0 & 0 & 1 & -1 \end{pmatrix}. \tag{17}$$

In the PAMM imaging experiment, the amplitudes of PA signals generated by six polarized lights are $PA_H$, $PA_V$, $PA_P$, $PA_M$, $PA_R$, $PA_L$, respectively. Then PA signal matrix can be expressed as

$$\mathbf{S_{PA}} = \begin{pmatrix} PA_H & PA_V & PA_P & PA_M & PA_R & PA_L \\ PA_H & -PA_V & 0 & 0 & 0 & 0 \\ 0 & 0 & -PA_P & -PA_M & 0 & 0 \\ 0 & 0 & 0 & 0 & PA_R & -PA_L \end{pmatrix}. \tag{18}$$

Therefore, the PAMM that carries characteristic information of the optically-anisotropic molecules can be obtained as a 4×4 matrix with 16 elements by $\mathbf{M_{PA}} = \mathbf{S_{PA}} (\mathbf{S_{Laser}})^{-1}$ as

$$\mathbf{M_{PA}} = \begin{pmatrix} M_{11} & M_{12} & M_{13} & M_{14} \\ M_{21} & M_{22} & M_{23} & M_{24} \\ M_{31} & M_{32} & M_{33} & M_{34} \\ M_{41} & M_{42} & M_{43} & M_{44} \end{pmatrix}. \tag{19}$$

As the PAMM imaging technique solely exploiting the anisotropic optical absorption information of the excited molecules, compared with the optical MMI, the proposed PAMM is more capable of providing physically explicit information. Eq. (19) showed that, the first line of $\mathbf{M_{PA}}$ indicates exact absorption information of the excited molecules under different polarized laser beams (indicated by Eq. (6)), which therefore is used to extract the information of microstructure, molecular interaction and molecular orientation.

**Experimental setup.**

To validate the proposed method, we established the PAMM experimental system, as shown in the Fig. 1D and Supplementary Fig. S3. A pulsed square wave laser (Model TECH-527 Basic, pulsed Q-switched lasers, Russia) with a 527 nm wavelength and a 1.0 kHz repetition frequency was used as the laser source. The laser beam is expanded and collimated through a spatial filtering system. After that, a polarizing beam splitter cubes (PBS, Thorlabs) is used to obtain a linearly polarized beam and then passed a quarter-wave plate ($1/4\lambda$ plate) to produce circularly polarized laser beams. Finally, the laser beam is transformed into linearly polarized light through a polarizer. During the experiment, four linearly polarized light with a 45° polarization difference was obtained by rotating the polarizer. The left- and right-handed circularly polarized lights was obtained by removing the polarizer and rotating the $1/4\lambda$ plate to 90°. The excitation laser beams were focused through the NA=0.1 of objective (OB) lens. The generated PA signals were received by an ultrasonic transducer (a center hole diameter of 6 mm, focus of 15.6 mm, and central frequency of 10 MHz with 95% bandwidth), then transferred to a 30 kHz – 600 MHz low noise amplifier (LNA-650, RF BAY) and finally collected by a data acquisition card (DAQ, M3i.4110). A photodiode (PD) is exploited to monitor the power. A computer was used for controlling the system and data processing. To gain precision of repetitive movement, we used two high precision stepper motors (LNR502/M, Thorlabs) with 1 μm step size and their associated drives (BSC202, Thorlabs).

**Data analysis.**

PA, DOA, OTDM and CD imaging parameters of PAMM image was calculated in the time domain, and its maximum amplitude projection was computed to construct an image. In Figs.3-6, to highlight the DOA, OTDM and CD, information is shown only at positions where the corresponding PA amplitudes are above a certain threshold value (2 times noise level). Note the analysis and image of CD data is performed on absolute values. The final images were further processed with a median filter to reduce salt-and-pepper noise.

**Animal studies.**

The characterization of the APP/PS1 double transgenic mice has been described previously, which is a chimeric mice/human amyloid precursor protein with a Swedish mutation (Mo/HuAPP695swe) and a mutant human Presenilin 1 protein (PS1-dE9) in central nervous system neurons *(55)*. The genotype was confirmed by polymerase chain reaction (PCR) analysis of tail biopsies. In the present study, we used APP/PS1 transgenic mice with pathological features of AD and their non-transgenic WT littermates as control. All of the experimental mice were of the C57BL/6 background, and the non-transgenic WT and APP/PS1 transgenic mice were paired from the litters and housed under the same living conditions.

The meat samples (tendon and beef) were purchased from a local supermarket.

The present study was performed following the Guide for the Care and Use of Laboratory Animals (Institute of Laboratory Animal Resources, Commission on Life Sciences, National Research Council). This study was approved by the Institutional Animal Care and Use Committee of our university (South China Normal University, Guangzhou, China).

**Immunohistochemistry.**

All mice were deeply anesthetized with sodium pentobarbital (50 mg/kg intraperitoneally), and mice were perfused transcardially with ice-cold 0.01M PBS (pH=7.4) the brain was fixed in 4% paraformaldehyde (pH =7.4) for 12 hours and dehydrated through the gradient of 15% and 30% sucrose solutions until sank at 4 °C. The brains were mounted in Optimal Cutting Temperature (OCT) compound (AO, USA), and sequential coronal brain sections (10 μm thick) were obtained and mounted on polylysine-coated slides. For immunostaining, the sections were permeabilized with 0.5% Triton X-100 in PBS for 30 min and blocked with 5% bovine serum albumin at room temperature for 1 h and subsequently incubated with appropriately diluted primary anti-Aβ antibody (Covance, SIG-39320, dilution: 1:400) overnight at 4 °C. After washing with TBS/0.025% Tween 20, the sections were incubated with diluted Alexa Fluor-conjugated secondary antibodies for 2 hours at room temperature. After washes, sections were mounted, and images were acquired with a laser scanning confocal microscope (Zeiss LSM 880). Thioflavin T (Th-T) staining for fibrillar Aβ plaques was performed by incubating slides in 0.5% Th-T, followed by rinsing in ethanol and distilled water. The green fluorescent protein-stained plaques were visualized using fluorescence microscopy.

**Acknowledgments**

**Funding:** National Natural Science Foundation of China (Nos. 12174125, 61627827, 61805085), Science and Technology Planning Project of Guangdong Province, China (Nos. 2015B020233016, 2018A030310519), the Guangzhou Science and technology plan project (No. 201904010321), the Science and Technology Program of Guangzhou (No. 2019050001). This research was supported in part by the Intramural Research Program of 635 the NIH, including NIBIB (No. 1ZIAEB000015).

**Author contributions:** Z. Z., Y. S., S. Y. and D. X. designed the experiment. Z. Z contributed to the system construction and performed the sample fabrications. Z. Z. and Y. S. performed the experiment and data analysis. Y. S., W. Z., and Z. Z. wrote the draft of the manuscript. S. Y. and D. X. supervised the project. Q. S. prepared immunofluorescence imaging. Z. Z., Y. S., S. Y., W. Z and D. X. were involved in discussions. All authors took part in the discussion and revision and approved the final copy of the manuscript.

**Competing interests:** The authors declare that they have no competing interests.

# Figures

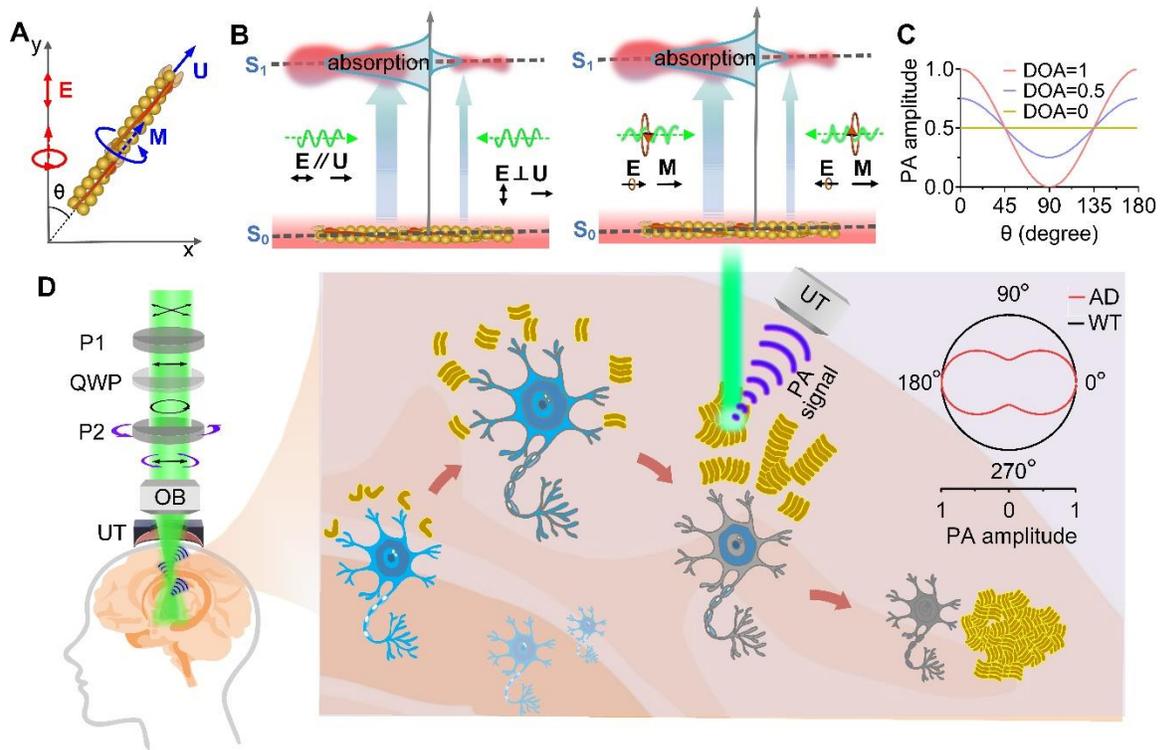

**Fig. 1 Schematic diagram of the photoacoustic Mueller matrix (PAMM) tomography.** (**A**) Schematic diagram of anisotropic absorption by a single chromophore. **E**, **U** and **M** are the electric vector of incident light (EVIL), the electric transition dipole moment (ETDM) and magnetic transition dipole moment (MTDM) of the chromophore, respectively. $\theta$ is the angle between **E** and **U**. (**B**) The transition diagram of a chromophore with dichroism and circular dichroism (CD) excited by linearly polarized light and circularly polarized light. (**C**) The relationship between photoacoustic (PA) amplitude and $\theta$ at different DOA values was studied. (**D**) Schematic diagram of experimental setup. And schematic diagram of PA signal generated by the polarized light excitation of Aβ with vector absorption properties in brain tissue of neurodegenerative diseases. P1, Polarizer-1; QWP, quarter-wave plate; P2, polarizer-2; OB, objective; UT, ultrasonic transducer.

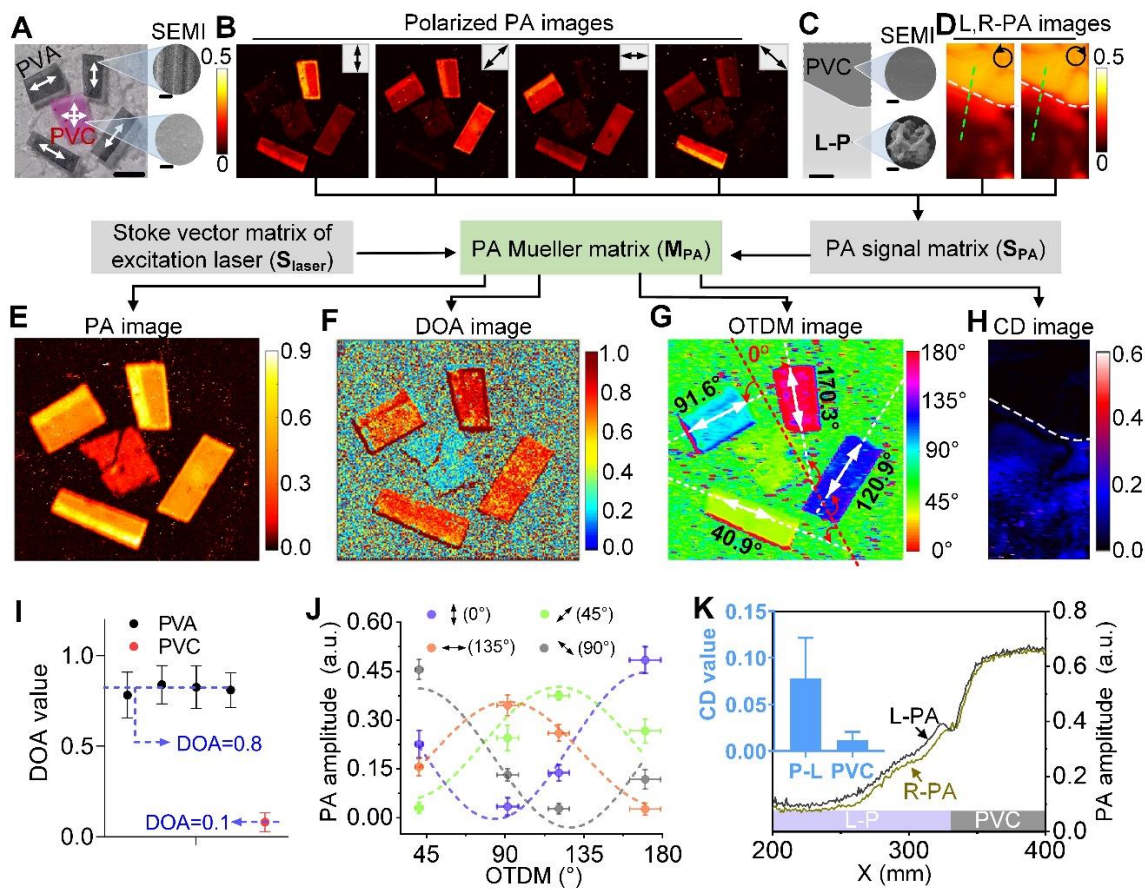

**Fig. 2 Demonstration of PAMM tomography capability.** (**A**) Sample photo. The white arrows indicate the ETDM orientation of the samples, with scale bar of 1 mm. The inset shows the scanning electron microscope images (SEMI) of polyvinyl alcohol (PVA) and polyvinyl chloride (PVC) respectively, with scale bars of 0.5 μm. (**B**) Polarized PA images of (**A**). The black arrows indicate the electric vector direction of excitation laser. (**C**) Schematic diagram of experimental sample, with scale bar of 1 mm. The inset shows the SEMI of L-polylactide (P-L) and PVC respectively, with scale cars of 0.5 μm. (**D**) The PA images under left- and right-handed circularly polarized lasers excitation for PVC and L-P samples. (**E**) PA image of the PVA and PVC samples that reflects the total absorption coefficient. (**F**) The degree of anisotropy (DOA) image that reflects the ordered degree of the samples. (**G**) The orientation of the transition dipole moment (OTDM) image that reflects the orientation of the chromophore. (**H**) CD image that reflects the chiral character of the samples. (**I**) The average statistics of five samples in (**F**) respectively. (**J**) The statistical results of PA signal amplitude in (**B**) as a function of the OTDMs in (**G**) for the four PVA samples. (**K**) Comparison of line profiles along the green dashed lines in the (**D**). The inset shows statistical results DOA value in (**H**).

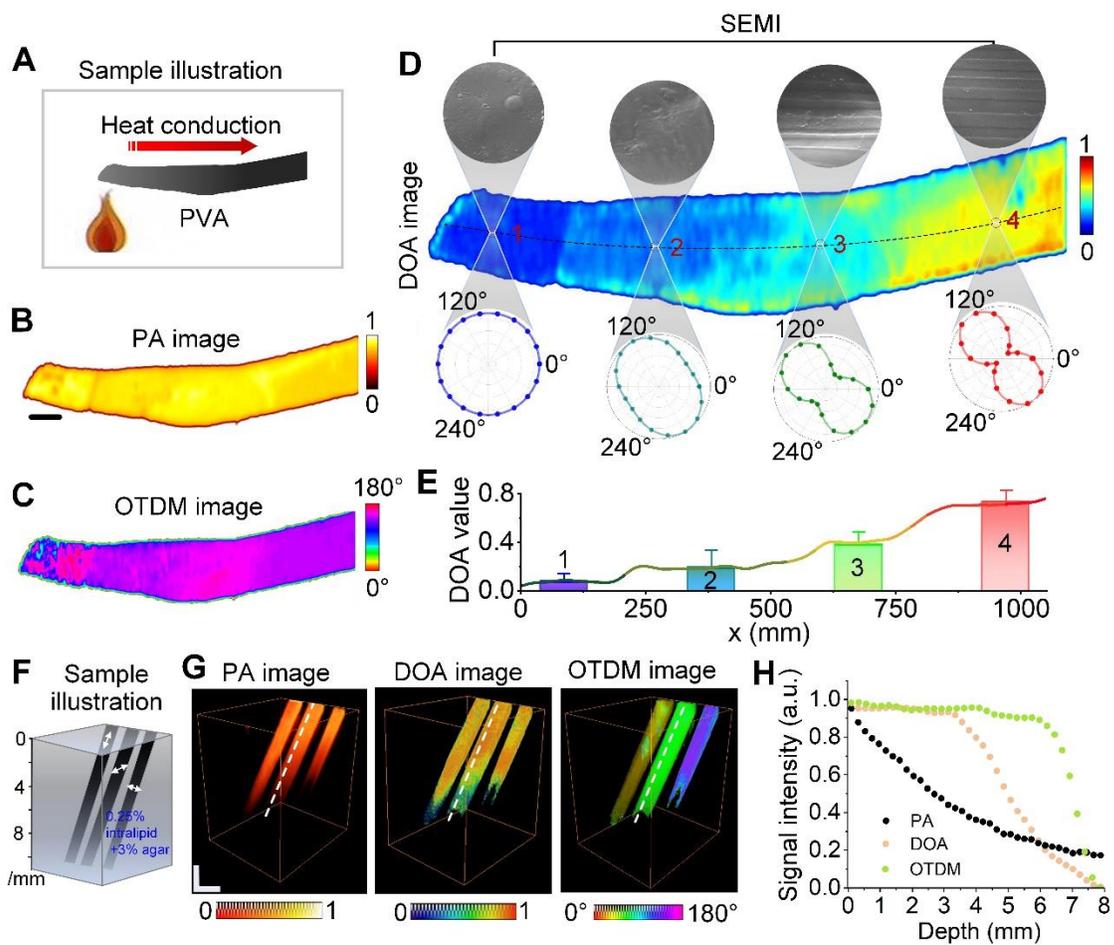

**Fig. 3 PAMM tomography application demonstration.** (**A**) Schematic diagram of a thermally damaged PVA sample. (**B**) and (**C**) are the obtained PA, OTDM imaging of the PVA sample, with scale bar of 1 mm. (**D**) DOA image. The inset shows the SEMIs of the sample and the variation of PA signal amplitude with angle $\theta$ in polar coordinates at the 1, 2, 3, 4 areas. (**E**) The average statistical histogram of DOA value for the four pointed areas and variation of DOA value along the black dashed lines in (**D**). (**F**) Schematic diagram of the experimental sample, simulating the scattering of light by biological tissues. White arrows indicate ETDM orientation of target. (**G**) Three-dimensional (3D) PAMM image of PVA samples. The corresponding PA image, DOA image, and OTDM image, with scale bar 2 mm. (**H**) Variation of PA, DOA, and OTDM signals intensity with probed depth along the white dashed lines in (**G**).

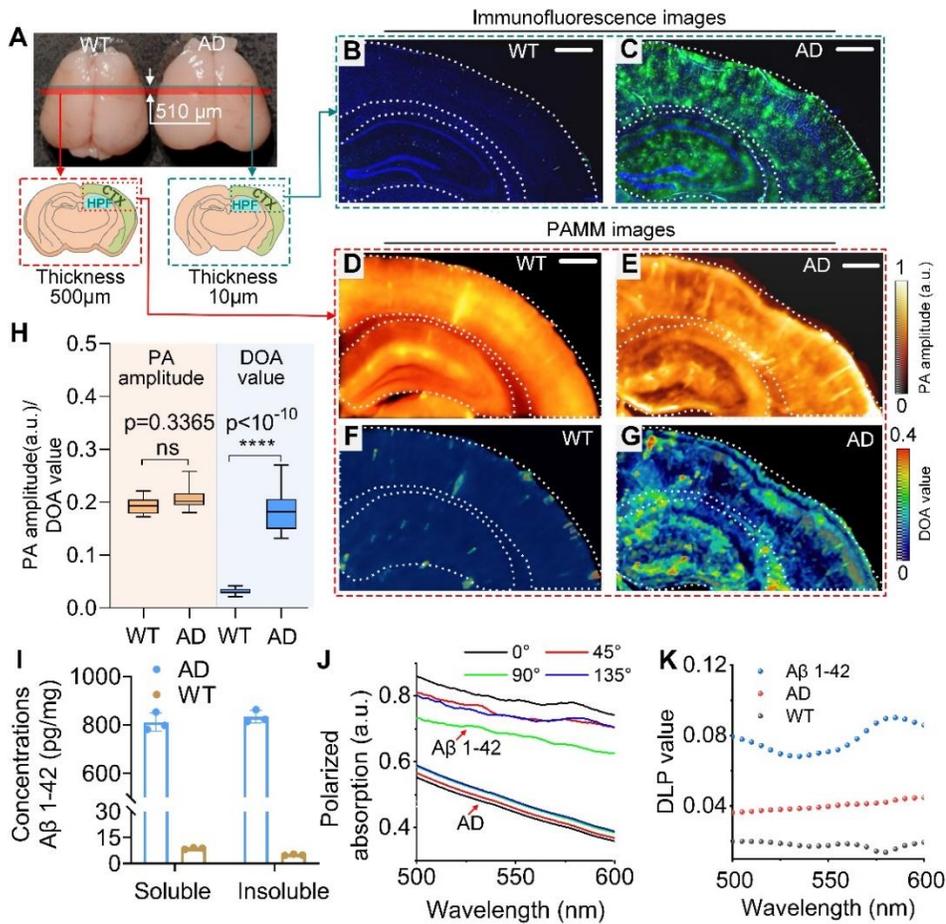

**Fig. 4 PAMM tomography imaging of APP/PS1 transgenic mouse (AD mouse) and non-transgenic wild-type mouse (WT mouse).** (**A**) Schematic diagram of 9-month-old AD and WT mouse brains. CTX, Cortical cortex; HPF, Hippocampal formation. (**B**) and (**C**) are immunofluorescence imaging of coronal brain sections (10 μm thick) of WT and AD mouse with Aβ-specific 6E10 antibody (green), respectively. The nucleus was stained with DAPI (blue). (**D-G**) PAMM images of coronal brain section (500 μm thick) of WT and AD mouse were obtained respectively (see Supplementary Videos 1, 2, 3 and 4.), with scale bars of 0.5 mm. (**H**) The box plots results of PAMM (**D-G**) images. For all boxes, the black central line represents the median, the square represents the mean. Statistical significance was tested using two-tailed Student's t-test. (**I**) The concentrations of Aβ 1-42 in the cortex and hippocampus were determined with an enzyme-linked immuno-sorbent assay (ELISA). (**J**) Polarized absorption spectroscopy of the concentration for 200 μM synthetic Aβ 1-42 fibrils and protein in the cortex and hippocampus of AD mouse. (**K**) Degree of linear polarization (DLP) value of Aβ 1-42 fibrils and protein in the cortex and hippocampus of AD mouse, WT mouse, respectively.

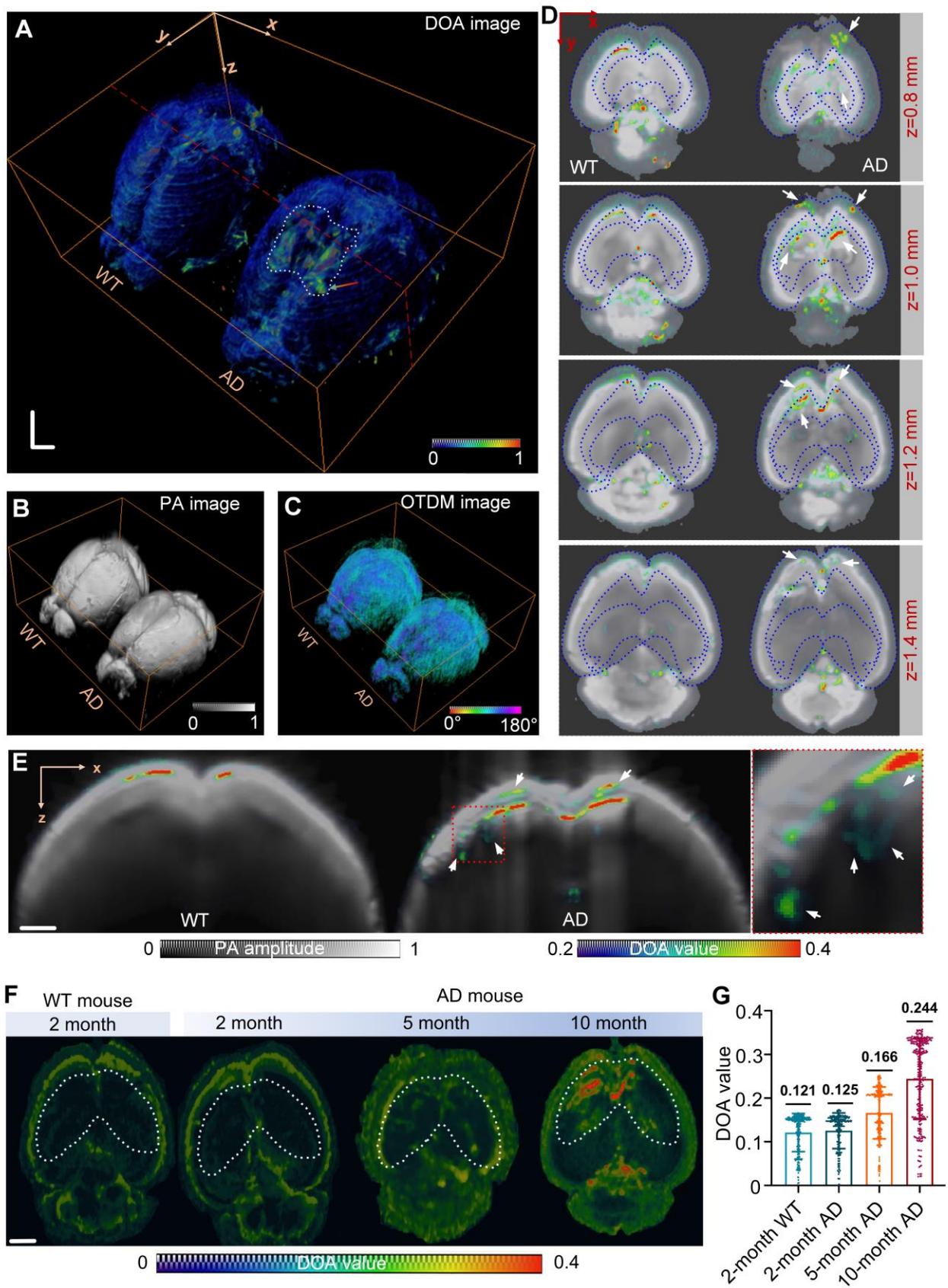

**Fig. 5 3D PAMM tomography imaging of AD mouse and WT mouse brain.** (**A**) 3D DOA image indicates larger DOA values in the white dashed circle of AD mouse than that of WT mouse. (**B**) 3D PA image. (**C**) 3D

OTDM image. (**D**) The overlay images of PA and DOA images at different imaging depth (see Supplementary Video 5.). (**E**) The overlay of PA images and DOA images at x-z cross-section (coronal section) along the red dashed line in (**A**). The inset shows enlarged image of the red box. (**F**) DOA images of AD and WT mice at different stages (see Supplementary Videos 6, 7and 8). (**G**) Histogram of the DOA values in the area of white dotted line of (F). The mean values of DOA are 0.121, 0.125, 0.166 and 0.244, respectively. All image scale bars are 1 mm.

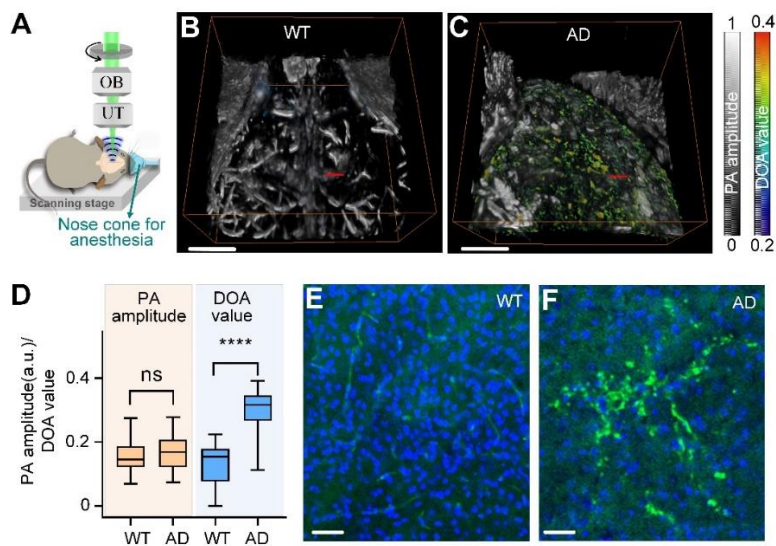

**Fig. 6 *In vivo* PAMM tomography image of AD mouse and WT mouse brain.** (**A**) Schematics diagram of the *in vivo* PAMM tomography image experiments. (**B**) The overlay images of PA and DOA images of WT mouse brain *in vivo*. (**C**) The overlay images of PA and DOA images of AD mouse brain *in vivo*. (**D**) The box plots results of PAMM (**B**, **C**) images in the area of brain tissue, ****P<0.0001. (**E**) and (**F**) are immunofluorescence imaging at x-z cross-sectional section (coronal section) along the red line in (**B**) and (**C**) with Aβ-specific 6E10 antibody (green) respectively. The nucleus was stained with DAPI (blue). The images (**B**, **C**) scale bars are 2 mm. The images (e, f) scale bars are 50μm.

# Supplementary information for
# Label-free visualization of amyloid plaques in Alzheimer's disease with polarization-sensitive photoacoustic Mueller matrix tomography

Zhenhui Zhang[1, 2*], Yujiao Shi[1, 2*], Qi Shen [1, 2], Zhixiong Wang[1, 2, 3], Da Xing [1, 2†], Sihua Yang[1, 2†]

1   MOE Key Laboratory of Laser Life Science & Institute of Laser Life Science, College of Biophotonics, South China Normal University, Guangzhou 510631, China.
2   Guangdong Provincial Key Laboratory of Laser Life Science, College of Biophotonics, South China Normal University, Guangzhou 510631, China.
3   Laboratory of Cellular Imaging and Macromolecular Biophysics, National Institute of Biomedical Imaging and Bioengineering, National Institutes of Health, Bethesda, Maryland 20892, USA

[*]*These authors contributed equally to this work.*
[†]*Corresponding author. Email: xingda@scnu.edu.cn; yangsh@scnu.edu.cn*

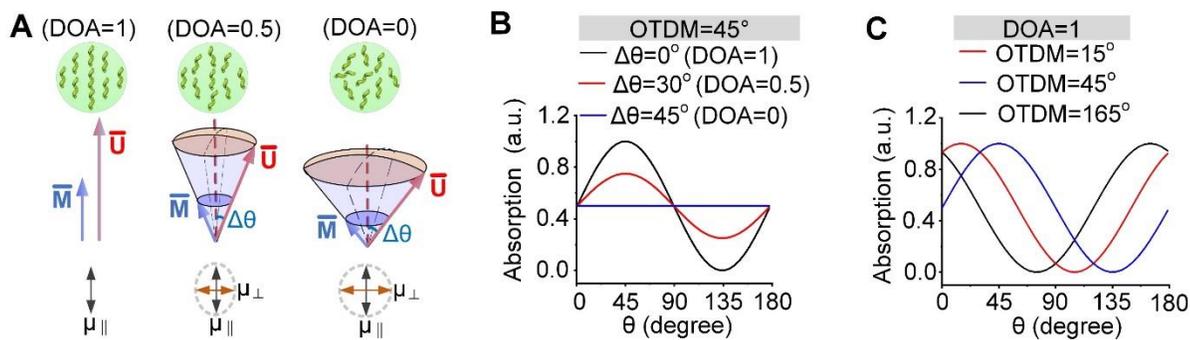

**Figure S1.** (**A**) When a group of chromophores is simultaneously excited in a laser spot, their average transition dipole moment ($\overline{U}, \overline{M}$) is distributed in the angle of $\theta+\Delta\theta$, where the degree of molecular arrangement can be reflected by the degree of anisotropy (DOA). $\Delta\theta$ is the dispersion angle. $\mu_{//}$ and $\mu_{\perp}$ are the spatial distributions of the absorption coefficient perpendicular and parallel to the optic axis, respectively. (**B**) When a group of chromophores is simultaneously excited in a laser spot, the relationship between the absorption intensity and the angle $\theta$ for different DOA values. (**C**) The relationship between absorption intensity and angle $\theta$ for different orientation of the transition dipole moment (OTDM) angles.

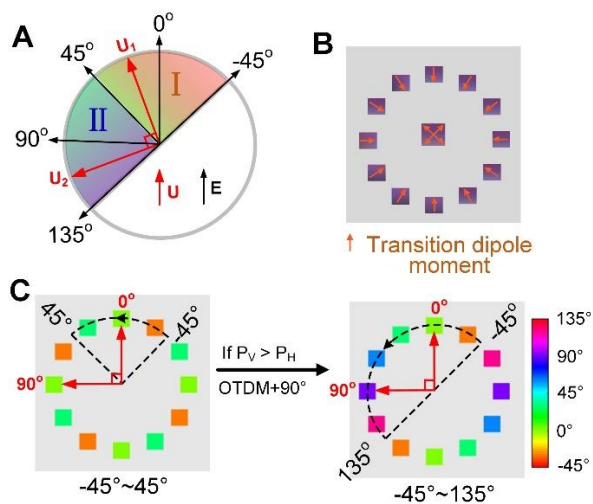

**Figure S2.** (**A**) Schematic diagram of OTDM imaging. **E**, and **U** are the electric vector of incident light and the ETDM, respectively. (**B**) Sketch map of the OTDMs for the virtual samples simulated by computer. The DOA value for these samples is set to be 1. (**C**) The simulated OTDM image of (**B**) according to Eq. (8).

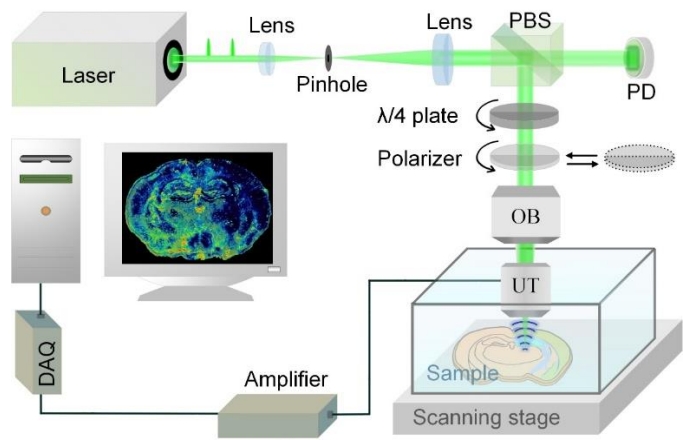

**Figure S3. Experimental setup for the PAMM tomography imaging.** Linear and circularly polarized light are produced by adjusting the polarizer. PBS, polarizing beam splitter cubes; OB, objective; PD, photodiode; UT, ultrasonic transducer; DAQ, data acquisition system.

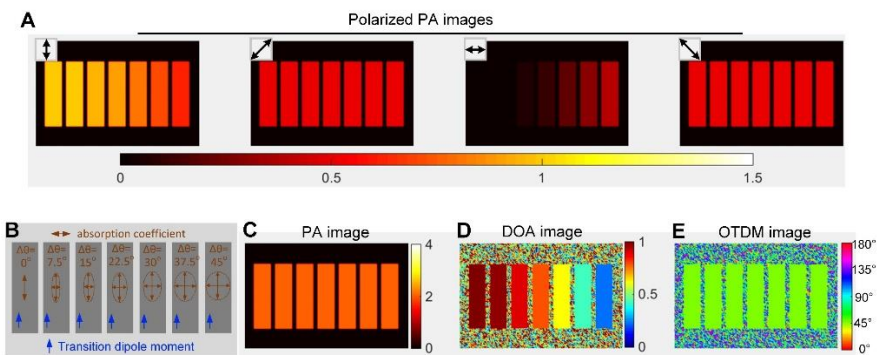

**Figure S4. Computer simulation of PAMM tomography imaging of virtual samples with different DOA values.** (**A**) Computer simulation of polarized PA images excited by linearly polarized light with different electric vector directions. The black arrows indicate the electric vector direction of excitation laser. (**B**) The schematic diagram of the virtual samples. (**C**), (**D**) and (**E**) are the obtained PA imaging, DOA imaging and OTDM imaging of the virtual samples by computer simulation according to the PAMM method, which are consistent with the setting parameters, proving the correctness of the method.

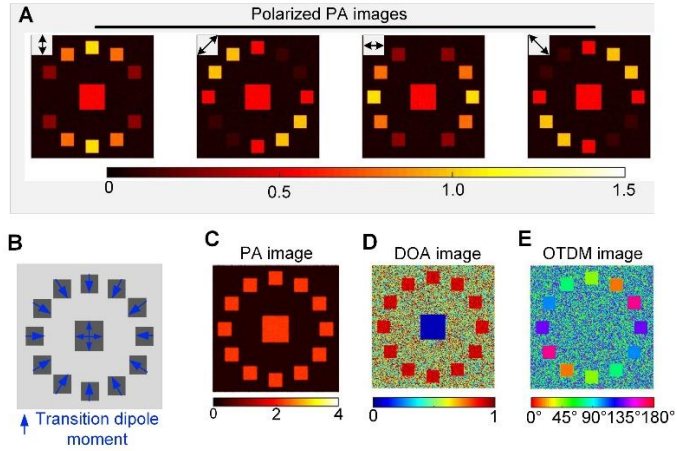

**Figure S5. Computer simulation of PAMM imaging of virtual samples with different OTDMs.** (**A**) Computer simulation of polarized PA images excited by linearly polarized light in different electric vector directions. (**B**) The schematic diagram of the virtual samples with different the OTDMs. The DOA value of the samples is set to be 1. (**C**), (**D**) and (**E**) are the obtained PA imaging, DOA imaging and OTDM imaging of the virtual samples by computer simulation according to the PAMM method.

The effectiveness of the proposed PAMM imaging is verified by computer simulation. Fig. S4B shows the virtual samples with different DOA values and certain OTDMs. The total optical absorption coefficient of the samples is set to be the same. Fig. S4A shows the obtained polarized PA images under the four linearly polarized laser with different electric vector directions of incident light. Fig. S4C shows the PA image with the same total optical absorption coefficient. Fig. S4D and Fig. S4e are the obtained DOA and OTDM images, respectively, which are consistent with the setting values in the computer simulation.

Furthermore, PAMM imaging of virtual samples with different OTDMs is simulated in Fig. S5, where the simulation results are in good agreement with the setting values in the computer simulation. These results verify the correctness of the proposed method for DOA and OTDM imaging.

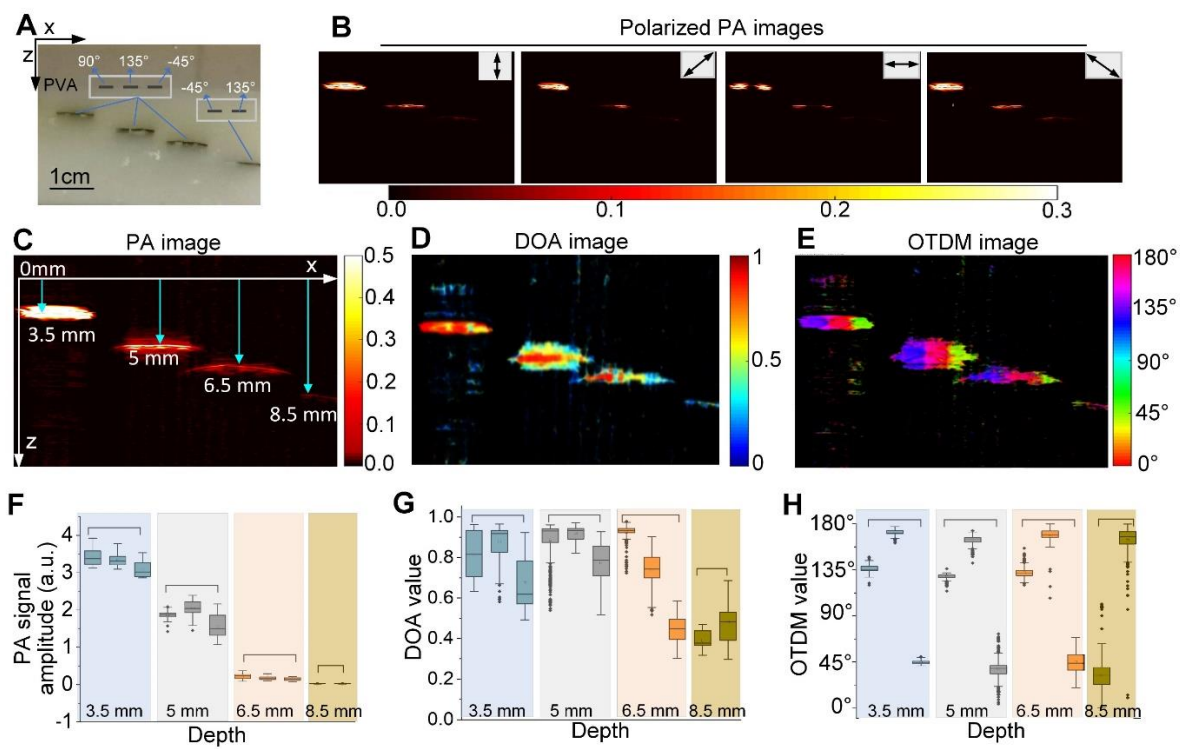

**Figure S6. PAMM Depth imaging**. (**A**) Sample photographs (0.25% intralipid and 3% agar in distilled water). **b**, Polarized PA images excited by linearly polarized light in different electric vector directions. (**C**), (**D**) and (**E**) are the obtained PA, DOA and OTDM images of the PVA sample by the PAMM method. (**F**) Statistical analysis of the PA signal amplitude for the samples with different depths in (**C**). Statistical analysis of the DOA values for the samples with different depths in (**D**). (**H**) Statistical analysis of the OTDMs for the samples with different depths in (**E**).

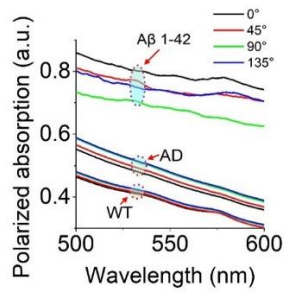

**Figure S7.**, Polarized absorption spectroscopy of the concentration for 200 μM synthetic Aβ 1-42 fibrils and protein in the cortex and hippocampus of AD mouse, WT mouse, respectively.

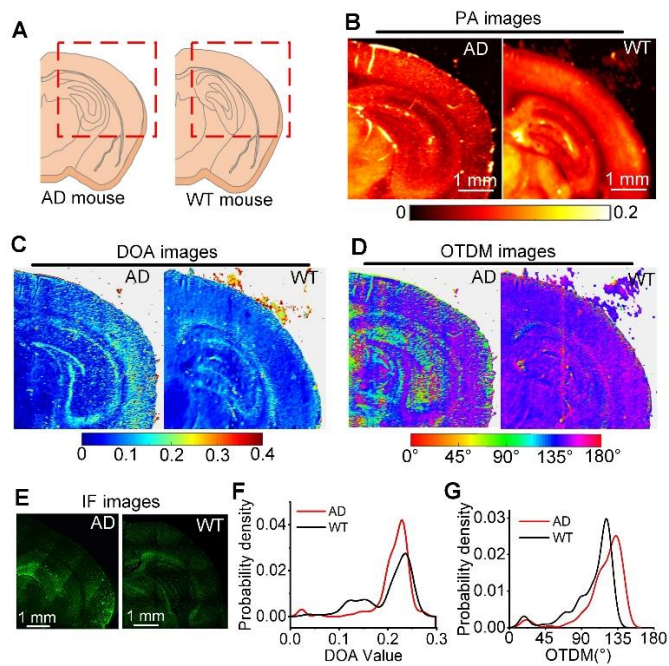

**Figure S8. PAMM imaging of 10-month-aged AD mouse and WT mouse brain.** (**A**) Sketch map of Sample. (**B**), (**C**) and (**D**) are the obtained PA, DOA and OTDM images by the PAMM method. (**E**) Immunofluorescence (IF) imaging after Aβ-specific 6E10 antibody immunostaining in brain slices of AD mouse and WT mouse. The IF imaging results of AD mouse showed that there were a lot of Aβ plaques in the cortex and hippocampus. (**F**) The statistical distribution of DOA values for AD mouse and WT mouse brain imaging results of (**C**). (**G**) The statistical distribution of OTDM values for AD mouse and WT mouse brain imaging results of (**D**).

In Fig. S8C, the value of DOA for AD mouse is higher, especially in the cerebral cortex, which is consistent with the hypothesis of Aβ deposition. The results of Fig. S8C and Fig. S8E are well corresponding. In addition, the distribution statistics of DOA values for AD mouse and WT mouse were analyzed in Fig. S8F, and it can be seen that the proportion of DOA values for AD mouse distributed in larger areas was higher. The distribution of OTDM values for AD mouse is obviously different from that for WT mouse, especially in the hippocampus. The distribution of OTDM value for AD mouse was more extensive than that for WT mouse (Fig. S8G). In general, this experiment well verified that PAMM imaging can distinguish the difference between AD mouse and WT mouse.

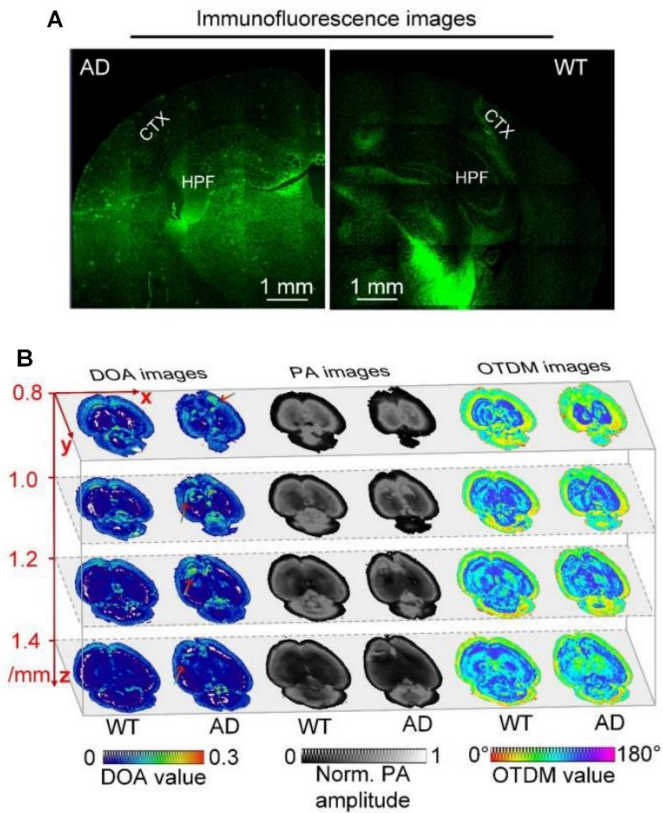

**Figure S9.** (**A**) IF imaging after Aβ-specific 6E10 antibody immunostaining in brain slices of AD mouse and WT mouse (corresponds to the Fig. 5E). (**B**) In the different depth x-y plane, DOA image, PA image and OTDM image. The DOA imaging can better distinguish the brain lesions of AD mouse and WT mouse.

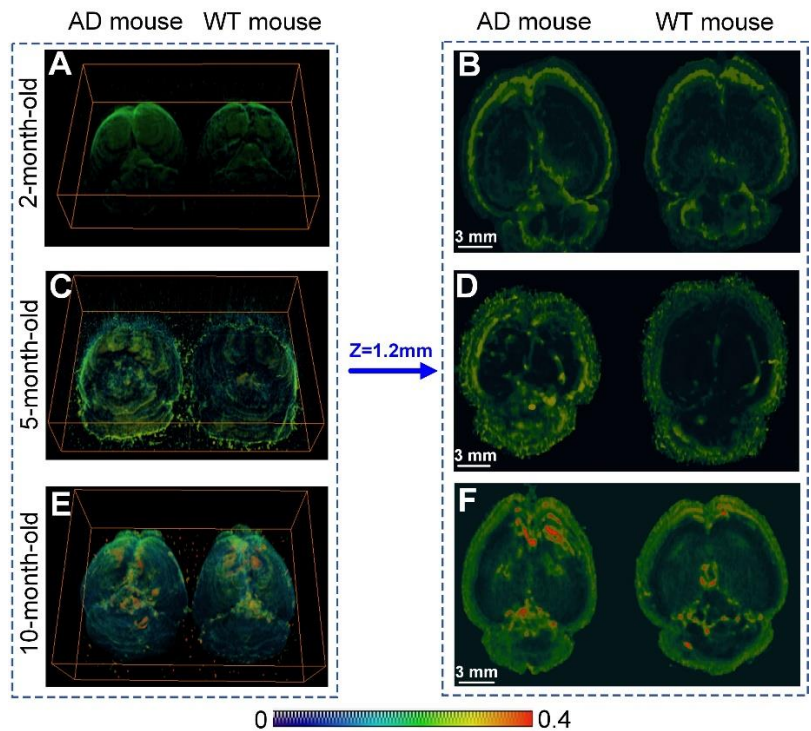

**Figure S10. DOA imaging of AD and WT mice brain at different stages.** (**A**) 3D DOA images of 2-mouth-old AD and WT mice brain. (**B**) x-y cross-sectional DOA images at z=1.2 mm depth of (**A**). (**C**) 3D DOA images of 5-mouth-old AD and WT mice brain. (**D**) x-y cross-sectional DOA images at z=1.2 mm depth of (**C**). (**E**) 3D DOA images of 10-mouth-old AD and WT mice brain. (**F**) x-y cross-sectional DOA images at z=1.2 mm depth of (**E**).

It can be seen that there is no significant difference between the 2-month-old AD and WT mice (Fig. S10A, B). However, there was a significant difference between the 5, 10-month-old AD and WT mice (Fig. S10C-F). Compared with WT group, the data of DOA value in AD group was higher, and the value of the DOA data increased with age.